# Technology of fabrication superconducting free-standing structures (FSS)

Tarkhov M.A., Mumlyakov A.M., Shibalov M.V., Porokhov N.V., Vovk N.A., Nekludova P.A., Trofimov I. V. and Filippov I.A.

Institute of Nanotechnologies of Microelectronics RAS, Nagatinskaya 16a-11, Moscow, Russia

## Abstract

In this study, a method for fabrication of superconducting microstructures that are partially or completely isolated from the substrate has been proposed. Two configurations of suspended micro-bridges have been suggested, i. e., the first structure that features a T-shaped etching of the substrate and the second structure which is completely separated from the substrate through periodically positioned supports. The creation of suspended structures is based on the principle of gas-phase etching of amorphous silicon oxide in a mixture of hydrogen fluoride (HF) and ethanol gases. In the course of the experiments, it has been discovered that suspending micro-structures in the configuration of a micro-bridge results in a slight reduction in superconducting characteristics, ranging from 10 to 15% of the initial parameters. It has also been demonstrated that the thermal coupling between the film and the substrate significantly affects the dissipation of thermal energy. The power dissipated into the substrate at room temperature can vary by up to 250 times based on the value of the micro-bridge undercutting.

## 1. Introduction

In contemporary microelectronics, various technologies and methodologies for fabricating structures that are detached from the substrate, such as MEMS and NEMS, are widely recognized and employed. However, the fabrication of structures that are isolated from the substrate while retaining superconducting properties presents a significant technological challenge, primarily due to the high susceptibility of superconducting materials to active media. It is also worth noting that modern superconductor-based devices are developed using ultrathin films. For instance, the operation of a superconducting single-photon detector is predicated on the principle of inducing local heating of a film section subsequent to the absorption of a single photon, thereby resulting in the formation of a hot spot [1,2]. The effectiveness of this process is closely linked to the thickness of the film. Thus, it requires the use of ultrathin films measuring only a few nanometers in thickness [3]. Besides, the detection efficiency, intrinsic noise, and performance of superconducting bolometers are influenced by the volume of the sensing element, which is directly correlated with the film thickness [4, 5]. Superconducting ultrathin films deposited on amorphous substrates often experience significant stress as a result of structural disparities between the film structure and the substrate, as well as the synthesis conditions involved in film formation. The property of ultrathin films and the potential separation from the substrate will inevitably lead to structural breaks.

One of the most effective materials utilized in contemporary cryoelectronics is niobium nitride. This material is favored due to its convenience in terms of synthesis and etching. Besides, niobium nitride exhibits high superconducting parameters, with a critical transition temperature of approximately 12-14 K [6, 7] and the high current density of up to 10 MA/cm2 [8] for films that are 5-7 nm thick. The most prevalent technique for depositing niobium nitride films involves magnetron sputtering of a niobium target in an argon atmosphere, with the presence of nitrogen or ammonia. Currently, the plasma-enhanced atomic layer deposition technique for creating high-quality ultrathin films of niobium nitride with superior structural and superconducting properties is being actively developed [9, 10, 11, 12]. Thus, the work by Shibalov et al. has successfully demonstrated the synthesis of an epitaxial film of niobium nitride on a sapphire substrate at a relatively low temperature. This holds significant promise for the development of cryoelectronic devices tailored for diverse functional applications.

In a comprehensive review of the existing literature regarding the development of superconducting structures detached from the substrate, no reliable technological method for achieving complete or partial isolation of superconducting structures from the substrate was identified. The majority of the content pertains to the methodology of suspending structures on a silicon nitride membrane, commonly referred to as spiderweb technology [13, 14, 15, 16, 17, 18]. It is worth noting that a considerable portion of the research is focused on high-temperature superconductors utilizing YBaCuO as a base. These studies demonstrate the fundamental potential for separating the film from the substrate [19].

The operational principles of superconducting nanowire single-photon detectors (SNSPD) should also be carefully considered. The detection of single photons relies on the principle of inducing local heating in a narrow superconducting stripe segment, located close to a critical current, following the absorption of a single photon [20]. The dynamics of the heated region are effectively characterized by the "hot spot" model [21] and align closely with experimental observations. The work by Yang et al. introduces an electrothermal relaxation model of SNSPD [22]. One of the relaxation channels for the hot spot is the transfer of nonequilibrium phonons into the substrate, contributing to its cooling. Therefore, it is scientifically justified and relevant to focus on controlling the cooling channel of a superconducting stripe through a substrate. It is evident from the hot spot model that regulating the cooling channel through the substrate will result in a considerable expansion of the hot spot diameter, consequently enhancing the detection capability.

In this paper, a novel technological approach is introduced for creating suspended superconducting structures of two distinct types. The first type involves a structure on a T-shaped substrate with a controlled undercutting parameter (uc), while the second type comprises a structure with complete isolation from the substrate, achieved through periodically positioned supports along the stripe. The studied structures consist of micro-bridge configurations with a width of 1 μm and a length of 50 μm.

## 2. Technology

The creation of suspended structures of both types is based on the principle of gas-phase etching of amorphous silicon oxide in a mixture of hydrogen fluoride (HF) and ethanol gases. An amorphous silicon oxide layer was deposited on a 4-inch PECVD silicon wafer using the silicon-containing monosilane precursor (SiH4), along with nitrous oxide (N2O) and nitrogen (N2). Before depositing, the wafer was processed in $N_2O$ plasma for cleaning from organic contamination. The substrate temperature was maintained at 350°C, while the pressure in the process chamber was set to 1,000 mTorr. The SiH4/N2O/N2 gas flow rates were equal to 10/750/200 sccm, respectively. The power of the RF source was 20 W. Under these conditions, the deposition rate was 70 nm/min. The thickness of the amorphous silicon oxide layer was 550 nm. Figure 1 illustrates a series of technological procedures for the formation of structures partially or fully isolated from the substrate.

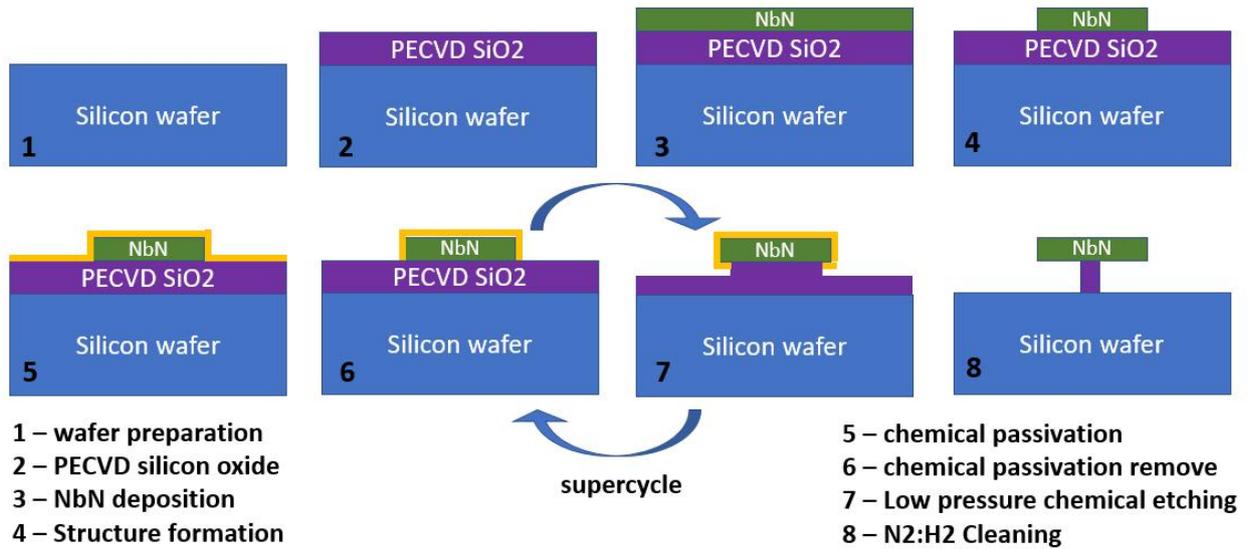

Figure 1. Technological flow processes of forming superconducting free-standing structure.

A thin film of NbN was deposited onto an amorphous silicon oxide substrate through the process of magnetron sputtering, utilizing a niobium target in a gas mixture consisting of Ar and N2 with a gas flow ratio of 27:8, respectively. The substrate temperature was 350°C, while the specific power of the magnetron was approximately 5 W/cm2. The film deposition rate was approximately 7 nm/min. The amorphous silicon oxide surface underwent treatment with Ar ions at an energy level of 1.5 keV prior to the deposition of niobium nitride in order to activate the surface.

Silicon oxide was etched isotropically at low pressure using a gas mixture comprising HF, ethanol ($C_2H_5OH$), and nitrogen ($N_2$). The inclusion of alcohol serves to inhibit the buildup of excess water on the wafer's surface. Furthermore, alcohol aids in the generation of chemically active ions $HF_2^-$ and the partial passivation of niobium nitride during the etching process of amorphous silicon oxide (refer to Figure 1 (5-6-7)). The pressure in the chamber was 75 Torr. Partial pressure of HF and $C_2H_5OH$ was 7 and 10 Torr, respectively. Following the etching of the amorphous silicon oxide, post-processing of the samples was conducted as ammonium hexafluorosilicate forms on the surface due to the presence of unbound nitrogen in the silicon oxide films. The residue was removed through plasma treatment using a 10:90 ratio of $H_2/N_2$ forming gas. The substrate temperature during processing in the forming gas was 200°C at a chamber pressure of 0.5 mTorr. The power of the RF source was 500W. In this instance, the decomposition of ammonium hexafluorosilicate occurs according to reaction (1), and its products are subsequently removed by a vacuum system.

$$(NH_4)_2SiF_6 \rightarrow NH_4HF_2 + SiF_4 + NH_3 \quad (1)$$

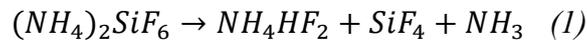

The micro-bridges under investigation were created through laser lithography, followed by plasma chemical etching of NbN using a gas mixture of SF6 and Ar. The ratio of gases was 15:25, respectively. The power of the RF and ICP sources was 40 W and 470 W, respectively, at a pressure of 8 mTorr in the process chamber. The termination areas were created from aluminum with a thickness of approximately 250 nm using the lift-off method. The width and length of the micro-bridges were 1 μm and 50 μm, respectively. In structures featuring regularly spaced supports, the local broadening of the stripe exhibited a characteristic size of 1.5 μm.

Due to the utilization of fluorinated plasma in NbN etching, there is a notable occurrence of parasitic etching on the substrate (amorphous SiO2), which can substantially impact the gas-phase etching profile.

To examine the etching profile of SiO2 while considering parasitic etching, this process was simulated. Figure 2a–2f (left part of Figure 2) display the initial profiles of parasitic etching. The initial etching depth values selected for the simulation are 0, 0.1, 0.2, 0.5, 0.9, and 1, which are contingent upon the thickness of the silicon oxide. A schematic representation in Figure 2 illustrates the profile of isotropic etching of amorphous silicon oxide through gas-phase etching. This process is contingent on the initial state of the substrate in relation to the film, particularly with regards to parasitic etching of the substrate during the formation of a micro-bridge.

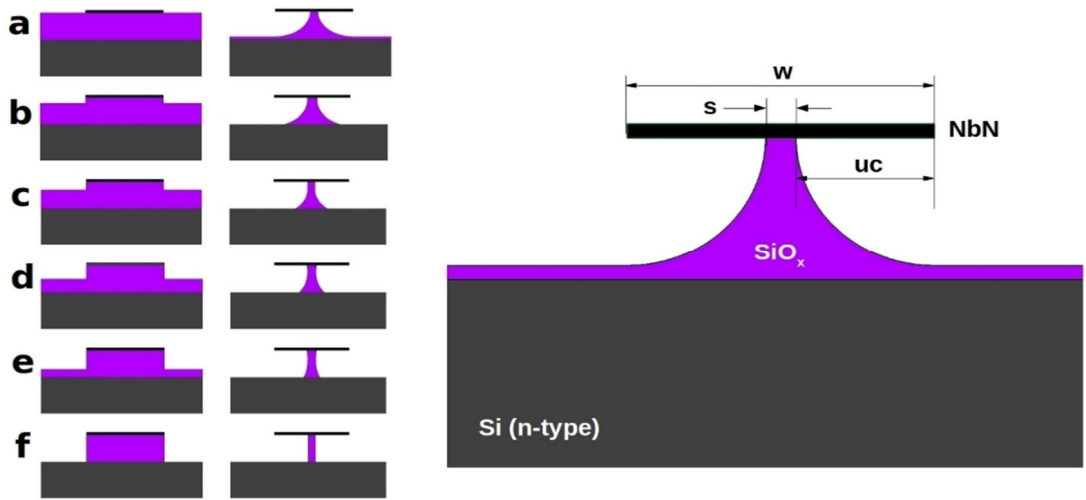

Figure 2. Schematic representation of the gas-phase etching profile of amorphous silicon oxide from the initial state of the substrate relative to the film.

The regulated technological parameter that defines the interaction between a superconducting stripe and a substrate is the undercutting, as illustrated in the right part of Figure 2. The coupling between the film and the substrate may be expressed as $s = w - 2uc$, according to a simple conversion formula.

The primary objective of this study is to advance the development of a technology capable of producing micron-scale structures with the capacity to regulate the insulation parameter from the substrate, and to examine the superconducting properties of these structures subsequent to the isotropic etching process.

Figure 3 illustrates scanning electron microscope (SEM) images of two types of suspended micro-bridges with distinct uc parameters. For micro bridges exhibiting a T-shaped etching profile, the uc parameter equals to 0 μm, 0.05 μm, 0.25 μm, 0.3 μm, 0.40 μm, 0.45 μm, as depicted in the upper section of Figure 3 (a–f). For structures that exhibit partial isolation from the substrate, the corresponding SEM image is presented in Figure 3 (a, b, c, d, e, f, g, h – medium part) with the respective parameters of 0 μm, 0.05 μm, 0.25 μm, 0.3 μm, 0.40 μm, 0.45 μm, 0.5 μm, and 0.55 μm. It can be observed that when the uc parameter exceeds 500 nm, the amorphous silicon oxide is entirely eliminated.

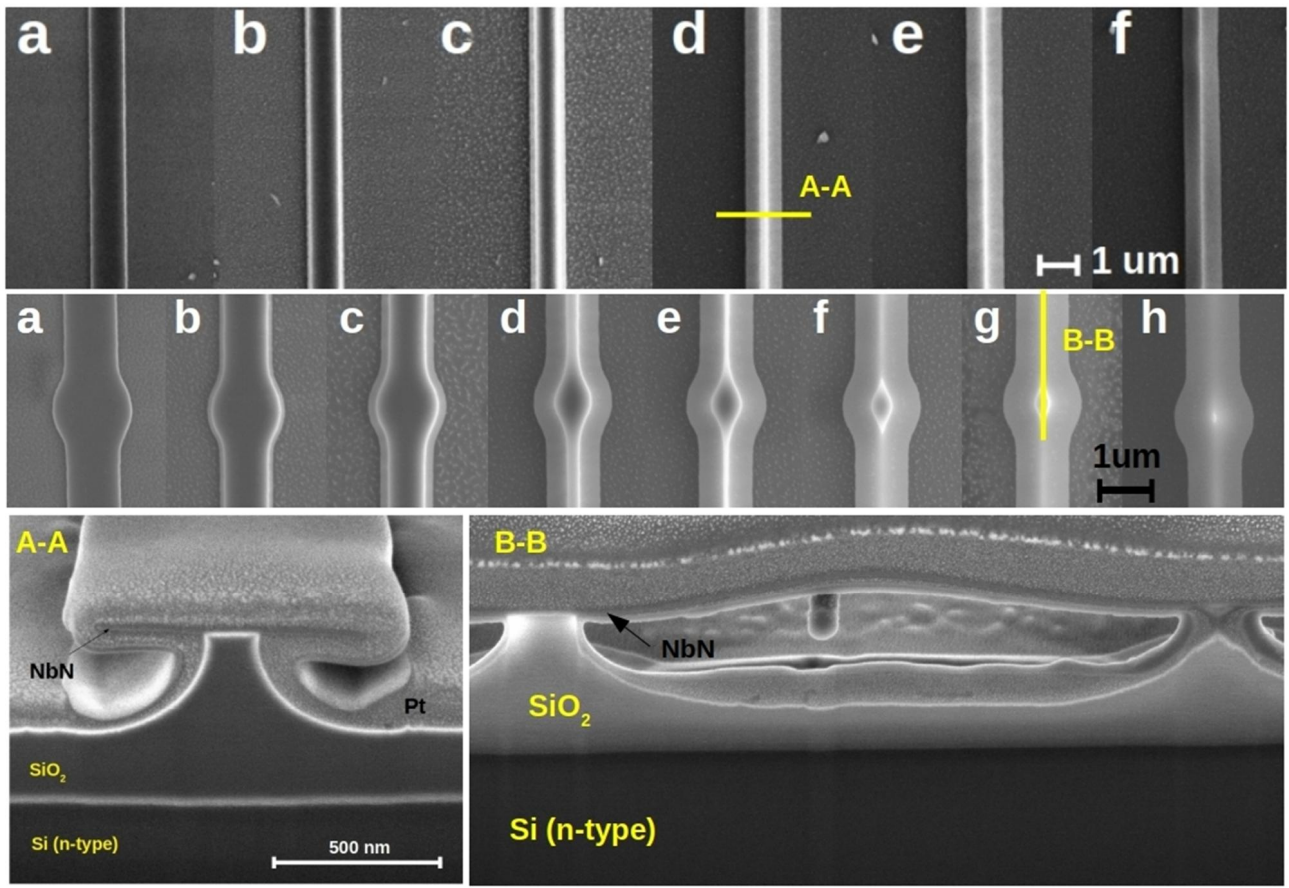

Figure 3. SEM image of suspended structures of a micro-bridge type with T-shaped support and freely suspended structures positioned on supports.

In order to assess the fabrication quality of the suspended structures, an examination of the etching profile of the samples under investigation was conducted. The focused ion beam (FIB) technique was employed to create a deep etching of the structure following the initial deposition of platinum from a metal-organic precursor. The etching profiles in various sections (A-A and B-B) are depicted in Figure 3, relative to the structure, with the functional layers labeled accordingly. In Figure 3 (A-A), a cross-section of a structure featuring a T-shaped micro-bridge etching profile with a uc parameter of 0.3 μm is depicted. It is evident that the film across the micro-bridge (section A-A) remains free from deformations. However, it experiences deformations along the stripe (section B-B). The anticipated outcome is typical for a polycrystalline film deposited on an amorphous substrate.

## 3. Results

In order to investigate the impact of the etching parameter (undercut parameter) on the electrophysical properties, two distinct microstructures were fabricated (refer to Figure 4 a, b) in accordance with potential and current contacts. Termination area were made of aluminum. In this study, an examination of electro-physical parameters was conducted within a closed-cycle cryostat utilizing an RDK-101D(L) cold head produced by SHI Cryogenics Group. The cryostat facilitated research across a broad temperature range, spanning from 300K to 2.3K. The Keithley 2460, a programmable low-noise precision source meter, served as the current/voltage source for the experiments. All measurements were conducted using the 4-point method.

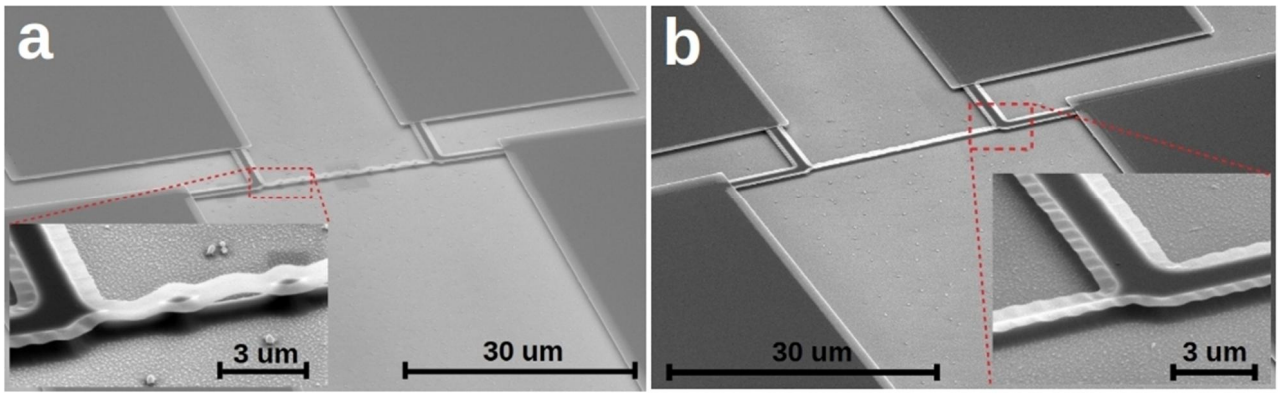

Figure 4. SEM image of two types of micro-bridges: the first type - a structure on a T-shaped support with a controlled undercutting parameter (uc), the second type - a structure with complete isolation from the substrate, achieved through periodically positioned supports along the stripe.

In Figure 5a, the current-voltage curve is depicted of samples with various values of the uc parameter, specifically for samples featuring localized widenings along the stripe (supported). The current-voltage curve was acquired under current stabilization mode at room temperature in a vacuum to eliminate the effects of convection cooling. The current variation occurred at a low (adiabatic) scanning velocity.

It is evident that for different uc parameters, the sample was burned out at varying bias currents (DC power). With the formula for calculating thermal power (P = IU), it is feasible to estimate the amount of Joule energy releasing in the stripe. In Figure 5b, two curves of normalized differential resistance from power are depicted for the initial sample and a sample with a uc parameter of 500 nm. In Figure 5c, the relationship between the power required for the structure being analyzed to burn out and the uc parameter is illustrated for samples exhibiting a T-shaped etching profile and partially isolated structures. Experimental evidence has demonstrated the significant role of the substrate in the cooling channel, showing a nearly two orders of magnitude impact.

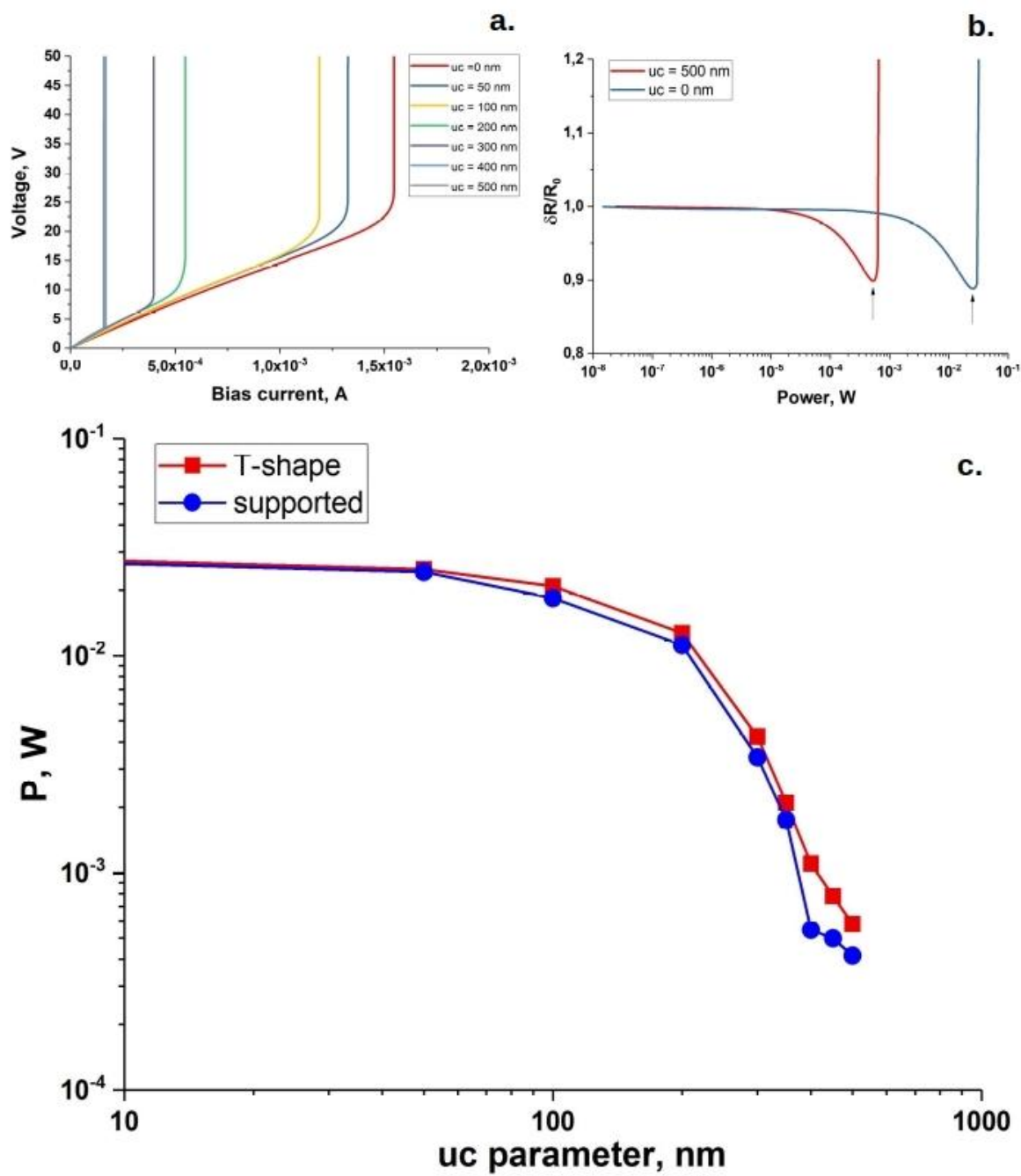

Figure 5. Characterization of structures isolated from the substrate at room temperature: a) Typical current-voltage curve for a sample with T-shaped support at various values of the uc parameter. b) Normalized differential resistance for the initial sample (uc = 0 nm) and at uc = 500 nm. c) Relationship between the thermal power at which the sample burns out and the uc parameter.

In the investigation of the superconducting properties of such structures, a notable observation of strong temperature hysteresis was made, seemingly attributable to the emergence of a thermal domain. In light of this, an experiment was carried out. The RT value measurements were conducted in a quasi-adiabatic mode, wherein the temperature varied at a sufficiently low rate and the sample's resistance remained constant, while the bias current was maintained at 0.1 of the critical current. The RT dependence was measured in two directions, from the superconductive to the resistive state (S-R), and vice versa (R-S). The heating and cooling dynamics is depicted in Figure 6. A typical heating/cooling rate was 1.8K/min.

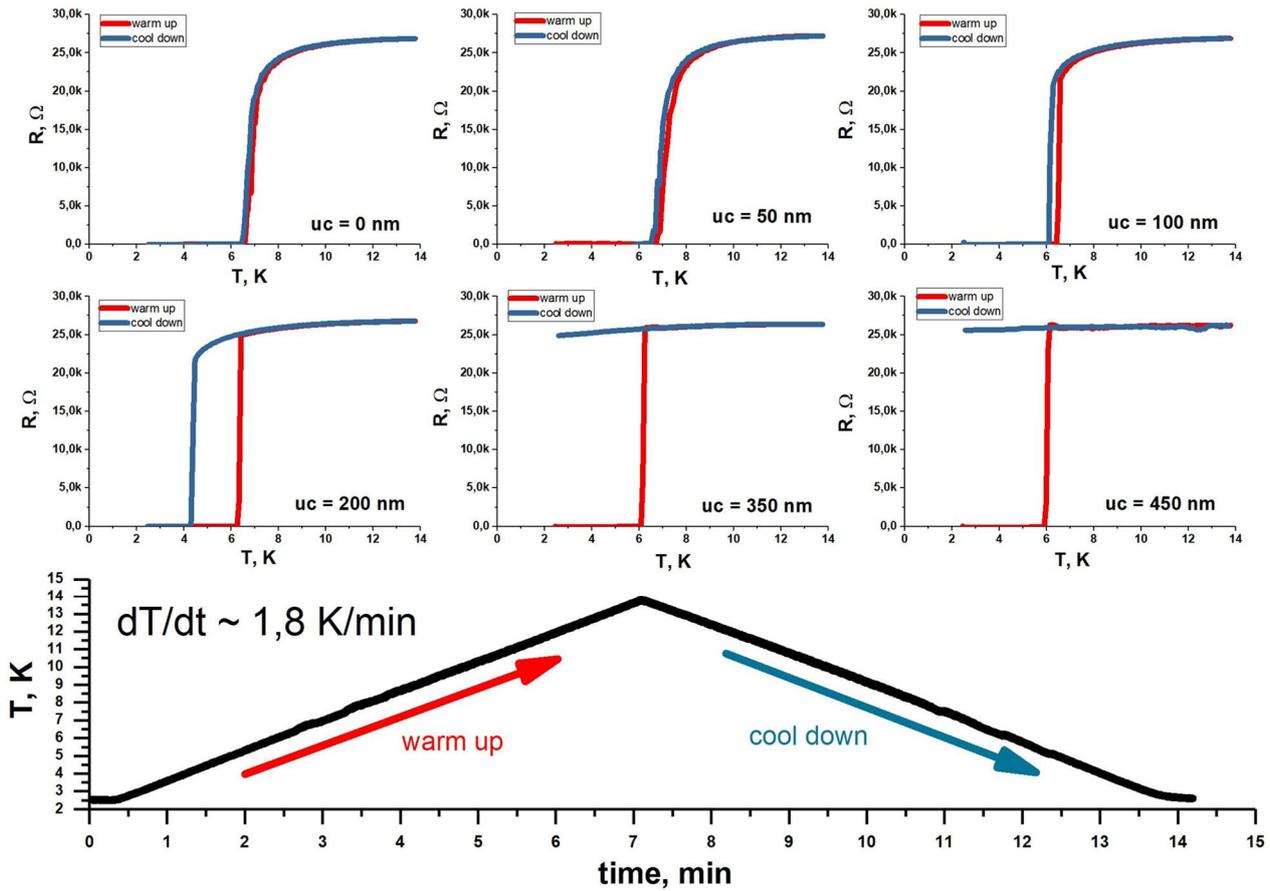

Figure 6. A set of RT dependencies measured for samples with varying IC parameters at a bias current of Ib/Ic ~ 0.1.

In Figure 6, the typical resistivity dependence of the temperature for micro-bridges that are partially isolated from the substrate is demonstrated. These results were obtained using various uc parameters with a fixed bias current of $0.1 \cdot I_c$. No transition to the superconducting state was observed at the reasonable time intervals for samples at uc = 350 nm and uc = 450 nm. The thermal power estimate, represented by $P_{DC}= I^2R$, released at the resistive area on the samples amounts to $P_{DC}$ = 0.266 mW. It is reasonable to assume that the energy expended on cooling the sample is lower than the dissipated power of $P_{DC}$. To transition the sample to a superconducting state, it is essential to reduce the bias current to zero.

In order to analyze the impact of the substrate etching process on niobium nitride, the current-voltage curve was measured at cryogenic temperatures, and the degradation of the critical current was subsequently examined. Figure 7 illustrates the relationship between the critical current and the uc parameter for the two types of structures. It is evident that during the substrate removal process, the critical current decreases by 10% and 15% based on the specific structure type. The observed effect is attributed to the presence of a chemically active medium consisting of HF and ethanol at relatively high processing temperatures.

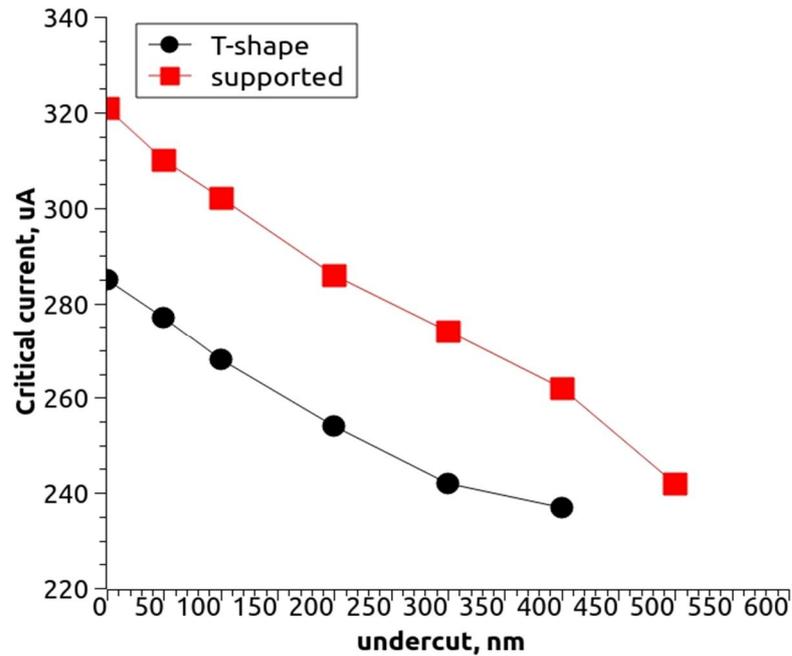

Figure 7. Dependence of critical current versus undercut parameter for two type samples.

The examination of structures near the critical current in the absence of light revealed triggers that are characteristic of dark counts in SNSPDs [23]. In the presence of light, a significant bolometric response is observed, leading to the latching states of such structures [24]. Considering this, an exhaustive examination of the bolometric and single-photon response was not conducted within the scope of this study.

## 4. Conclusion

The drive behind the advancement of a set of technological techniques for structures that are entirely or partially detached from the substrate is to establish innovative superconducting devices, enabling the manipulation of superconductivity through a mechanical degree of freedom. The technology enables the controlled manipulation of the mechanical properties, such as internal stresses in the film, which can lead to the emergence of unique superconducting structures.

Besides, the implementation of the proposed technology will undeniably have a substantial impact on the sensitivity parameters of superconducting bolometers to hot electrons. We hypothesize that as a result of the superconductor being isolated from the substrate, the dynamics of hot spot formation in the case of SNSPDs will deviate significantly from the classical representation.

## Data availability statement

All data that support the findings of this study are included within the article (and any supplementary files).

## Acknowledgments
The authors are grateful to Grant No. 122040800157-8 in support of the Ministry of Science and Higher Education of the Russian Federation. Fabrication and characterization were carried out at large-scale facility complexes for heterogeneous integration technologies and silicon + carbon nanotechnologies.